\documentclass[conference]{IEEEtran}
\IEEEoverridecommandlockouts

\usepackage[utf8]{inputenc}
\usepackage[T1]{fontenc}
\usepackage{cite}
\usepackage{amsmath,amssymb,amsfonts}
\usepackage{algorithmic}
\usepackage{graphicx}
\usepackage{xcolor}
\usepackage{booktabs}
\usepackage{multirow}
\usepackage{tabularx}
\usepackage{siunitx}
\usepackage{listings}
\usepackage[caption=false,font=footnotesize]{subfig}

\usepackage{url}      
\usepackage{hyperref}

\usepackage[acronym]{glossaries}
\makeglossaries

\def\BibTeX{{\rm B\kern-.05em{\sc i\kern-.025em b}\kern-.08em
    T\kern-.1667em\lower.7ex\hbox{E}\kern-.125emX}}

\newacronym{ba}{BA}{Biological Age}
\newacronym{ca}{CA}{Chronological Age}
\newacronym{hpp}{HPP}{Human Phenotype Project}
\newacronym{ml}{ML}{Machine Learning}
\newacronym{lgbm}{LightGBM}{Light Gradient Boosting Machine}
\newacronym{shap}{SHAP}{SHapley Additive exPlanations}
\newacronym{mlp}{MLP}{Multi-Layer Perceptron}
\newacronym{imt}{IMT}{Intima-Media Thickness}
\newacronym{bp}{BP}{Blood Pressure}
\newacronym{bmi}{BMI}{Body Mass Index}
\newacronym{hba1c}{HbA1c}{Hemoglobin A1c}

\begin{document}

\title{Few-Label Multimodal Modeling of SNP Variants and ECG Phenotypes Using Large Language Models for Cardiovascular Risk Stratification}

\author{
\IEEEauthorblockN{Niranjana Arun Menon\IEEEauthorrefmark{1},
Yulong Li\IEEEauthorrefmark{3},
Iqra Farooq\IEEEauthorrefmark{2},
Sara Ahmed\IEEEauthorrefmark{2},
Muhammad Awais\IEEEauthorrefmark{2},
Imran Razzak\IEEEauthorrefmark{3}}
\IEEEauthorblockA{\IEEEauthorrefmark{1}University of New South Wales, Australia}
\IEEEauthorblockA{\IEEEauthorrefmark{2}University of Surrey, United Kingdom}
\IEEEauthorblockA{\IEEEauthorrefmark{3}Mohamed bin Zayed University of Artificial Intelligence, UAE}
}
\maketitle

\begin{abstract}

Cardiovascular disease (CVD) risk stratification remains a major challenge due to its multifactorial nature and limited availability of high-quality labeled datasets. While genomic and electrophysiological data such as SNP variants and ECG phenotypes are increasingly accessible, effectively integrating these modalities in low-label settings is non-trivial. This challenge arises from the scarcity of well-annotated multimodal datasets and the high dimensionality of biological signals, which limit the effectiveness of conventional supervised models. To address this, we present a few-label multimodal framework that leverages large language models (LLMs) to combine genetic and electrophysiological information for cardiovascular risk stratification. Our approach incorporates a pseudo-label refinement strategy to adaptively distill high-confidence labels from weakly supervised predictions, enabling robust model fine-tuning with only a small set of ground-truth annotations. To enhance the interpretability, we frame the task as a Chain of Thought (CoT) reasoning problem, prompting the model to produce clinically relevant rationales alongside predictions. Experimental results demonstrate that the integration of multimodal inputs, few-label supervision, and CoT reasoning improves robustness and generalizability across diverse patient profiles. Experimental results using multimodal SNP variants and ECG-derived features demonstrated comparable performance to models trained on the full dataset, underscoring the promise of LLM-based few-label multimodal modeling for advancing personalized cardiovascular care. 
\end{abstract}


\section{Introduction}
Cardiovascular disease remains the leading global killer (about 20.5M deaths in 2023), making early risk stratification essential~\cite{Pcr_2023}. 
Early and accurate stratification of at-risk patients is essential for timely interventions and effective disease management. 
Traditional risk assessment methods, including genome-wide association studies (GWAS)~\cite{noauthor_genomewise_2025} and polygenic risk scores (PRS), have identified numerous single nucleotide polymorphisms (SNPs) associated with CVDs, many in non-coding regions influencing transcription factor (TF) binding and gene regulation~\cite{maurano2012systematic}. 
In parallel, electrocardiogram (ECG) phenotypes provide dynamic, physiologically relevant information about cardiac function. While both modalities independently contribute to risk assessment, their integration promises a richer understanding of genotype–phenotype relationships and improved predictive power. 

\begin{figure}
    \centering
    \includegraphics[width=0.9\linewidth]{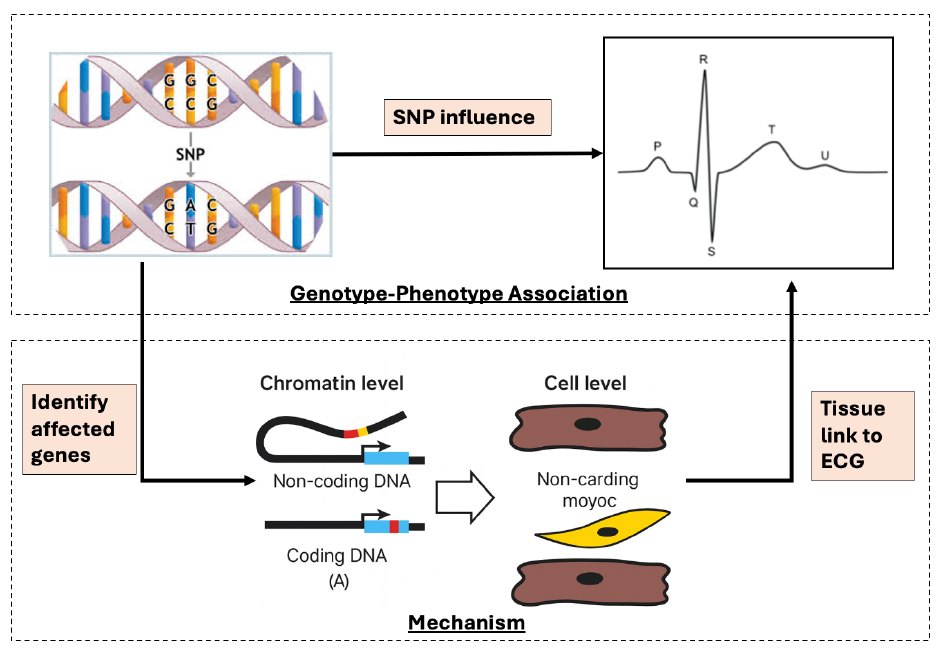}
    \caption{SNPs influence ECG intervals (PR, QRS, QT) across biological levels: (A) Chromatin—identify genes affected by coding or non-coding SNPs; (B) Cell—determine cell types expressing the gene; (C) Tissue—link gene expression to tissue electrophysiology and ECG phenotypes.}
    \label{fig:snps}
\end{figure}

Recent advances in multimodal AI modeling, such as the M-REGLE framework~\cite{zhou2025applying}, have demonstrated that combining genomic and electrophysiological data significantly enhances cardiovascular risk prediction. By jointly encoding ECG waveforms and genetic variants into shared latent representations, these models capture complementary biological signals that would otherwise remain obscured in unimodal analyses, thereby strengthening genotype–phenotype linkage discovery and improving polygenic risk estimation. However, a major bottleneck arises from the scarcity of high-quality labeled datasets that simultaneously contain genomic and electrophysiological annotations, limiting the usage of conventional fully supervised models. This scarcity of well-annotated multimodal datasets poses a major bottleneck for training deep learning models that rely on large quantities of labeled data. In many real-world biobank and clinical contexts, labels such as cardiovascular outcomes or expert-verified diagnoses are available only for a small subset of participants. Consequently, few-label and semi-supervised strategies are increasingly important for leveraging the vast amounts of unlabeled genotypic and electrophysiological data available.

However, recent advances in multimodal machine learning provide opportunities to overcome this challenge by enabling models to learn from heterogeneous data sources, even under limited supervision. For example, Gu et al. applied zero-shot and few-shot prompting with general-purpose LLMs such as GPT-3.5 and GPT-4 to automate clinical concept recognition, temporal relation extraction, and patient outcome prediction across large-scale EHR datasets~\cite{gu_scalable_2025}. Few-label learning strategies, including semi-supervised approaches, pseudo-labeling, and transfer learning from pre-trained foundation models, allow effective utilization of sparse labeled data while leveraging abundant unlabeled or weakly labeled information~\cite{srinivasan_applications_2025}~\cite{cakal_cardiovascular_2023}. Such strategies are particularly relevant in biomedical domains, where obtaining curated, high-confidence labels is often costly and time-consuming. By integrating SNP genotypes with ECG-derived features in a unified framework, few-label multimodal models can capture complex interactions that are difficult to identify through conventional statistical methods or single-modality analyses~\cite{sciencedirectFromZero}.

Large language models (LLMs), originally designed for natural language tasks, have recently demonstrated strong capabilities in representing structured and unstructured biomedical data~\cite{springerLargeLanguage}. LLMs have the ability to encode long-range dependencies, learn latent patterns, and integrate multimodal inputs, which is what makes them well-suited to few-label cardiovascular applications. Specifically, LLMs can embed SNP variants and ECG features into a shared latent space, enabling downstream classifiers to operate effectively even when labeled annotations are limited. Furthermore, Chain of Thought (CoT) reasoning can enhance interpretability, allowing the model to produce clinically meaningful rationales alongside risk stratification outcomes~\cite{sciencedirectComparativeEvaluation}~\cite{Zhang_Zhu_Ma_Xiong_Kim_Murai_Liu_27AD}. This combination of label-efficient learning, multi-modal integration, and interpretable reasoning provides a scalable framework for robust cardiovascular risk assessment across diverse patient cohorts.

In this study, we investigate the utility of few-label multimodal modeling of SNP variants and ECG phenotypes using large language models for cardiovascular risk stratification. Here, we present a framework that integrates sparse genetic and physiological features into a unified representation space, explores strategies for learning from limited labels, and evaluates predictive performance across cardiovascular risk outcomes. By addressing the challenges of multimodal integration under data scarcity, our work contributes to advancing robust, generalizable, and clinically relevant AI methods for cardiovascular health. Our key contributions include:

\begin{itemize}
    \item  Presenting a few labels chain of thought framework combining SNP variants and ECG-derived features using LLMs optimized for few-label learning that overcome the pseudo-label noise through top-$k$ selection method among the clusters to ensure the quality of data points, to distill high-confidence labels from limited annotated datasets, reducing noise and enhancing generalization
    \item Enable models not only to perform accurate risk stratification but also to generate clinically meaningful explanations that elucidate the underlying factors driving individual predictions. This capability enhances interpretability and supports clinical decision-making by providing transparent insights into how multimodal genetic and physiological features contribute to patient-specific cardiovascular risk.
    
    \item Experiment on multimodal dataset demonstrate improved performance and generalizability over traditional fully supervised or single-modality approaches, highlighting clinical applicability in real-world low-label scenarios.
\end{itemize}

\section{Data Sources and Integration}

\begin{figure*}[ht]
    \centering
    \includegraphics[width=0.760\textwidth]{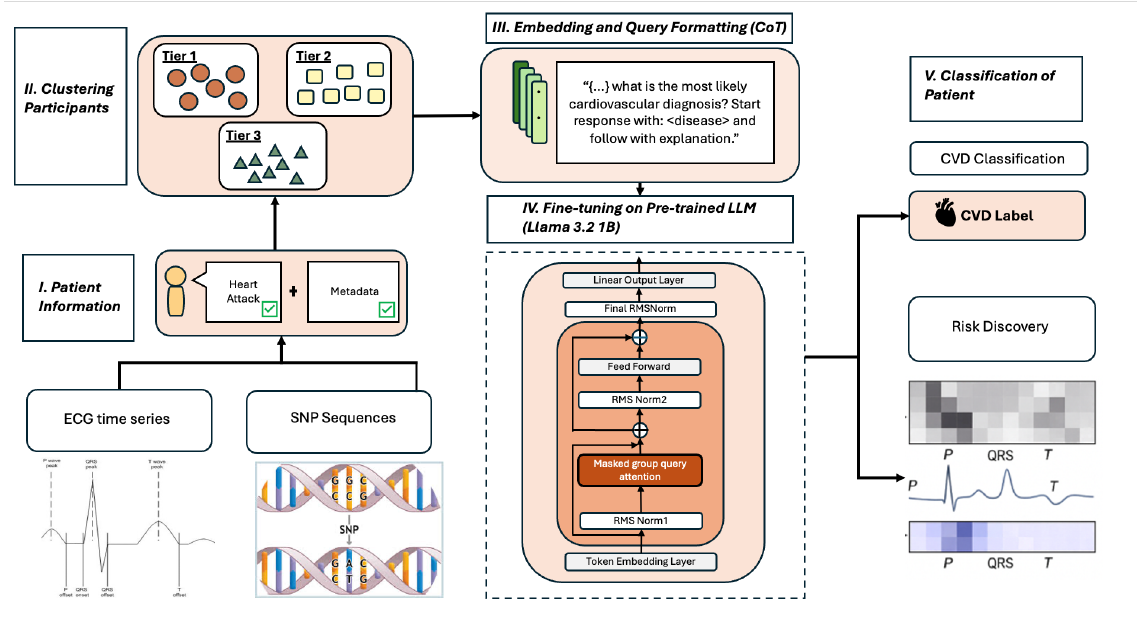}
    \caption{Linking SNP Variants and ECG Phenotypes for Explainable CVD Prediction.}
    \label{fig:example-generation}
\end{figure*}

Figure~\ref{fig:snps} illustrates the relationship between cardiac gene expression and electrophysiological phenotypes derived from electrocardiogram (ECG) signals. To investigate these relationships, we constructed a harmonized cardiogenomic dataset which integrates high-resolution single nucleotide polymorphism (SNP) genotyping with morphological and temporal ECG features. The dataset was curated from the PhenoAI HPP repository~\cite{phenoPhenoAI}, which includes cohort data from individuals residing in a specific geographic region (Asia/Jerusalem time zone), reducing potential temporal and environmental confounders related to circadian rhythm variations. This multimodal dataset provides a unified resource for modeling genetic variation alongside electrophysiological signatures to enhance cardiovascular risk stratification. The data collected was then preprocessed and standardized as follows:

\begin{itemize}
\item SNP variants and ECG-derived metrics underwent rigorous quality control, including filtering for minor allele frequency ($MAF < 0.01$), Hardy-Weinberg equilibrium (HWE), and genotype posterior (GP) thresholds. Following this, participants’ ECG-derived features were temporally aligned with corresponding genotype records using unique participant identifiers, ensuring consistent multimodal mapping. To maintain coherence across modalities, features were normalized and concatenated into joint embeddings, forming the foundation for multimodal fusion during training. The processed data were stored in \texttt{.json} format for efficient participant-level retrieval and processing. 

\item Clinical annotations of cardiac diagnoses, extracted from participants’ medical histories, were stored in \texttt{.csv} format to serve as phenotypic ground truth.
\end{itemize}

Datasets were linked using unique participant identifiers to maintain alignment across genomic, electrophysiological, and clinical domains. For multimodal modeling, we fine-tuned open-source large language models (LLMs) on tokenized biological sequences and dense ECG embeddings to create a shared representation space. Participants were stratified into three tiers based on label availability, ranging from fully supervised genotype-phenotype associations to cases requiring latent inference using only ECG morphology. This tiered structure allows us to evaluate model generalization across different levels of clinical supervision and biological abstraction. Fine-tuning was performed using Low-Rank Adaptation (LoRA)\cite{hu_2021_lora}, applied selectively to attention and MLP layers of the transformer architecture, with \texttt{rank = 8} and \texttt{alpha = 16}. Training was conducted on an AWS EC2 instance equipped with NVIDIA A10 GPUs using mixed precision (fp16) to improve memory efficiency and training speed. Model checkpoints were saved every 50 steps, and early stopping was employed based on the best validation accuracy. The overall fine-tuning workflow is summarized in Figure\ref{fig:example-generation}. This integration strategy preserves both genomic context and electrophysiological temporal structure, which is critical for capturing genotype–phenotype interactions relevant to cardiovascular risk.


\section{Feature Engineering and Pipeline Overview}
\subsection{Data Format and Preprocessing}

Our dataset includes $8,856$ participants with multimodal inputs in \texttt{.json} files containing ECG and SNP features, alongside a \texttt{.csv} file containing clinical condition labels. 
\begin{table*}[ht]
\centering
\caption{Cohort distribution and SNP / ECG feature counts before and after QC. Per-tier "unique variants observed" are approximated and computed proportionally by sample counts (variants are shared across tiers and thus rows are not additive).}
\label{tab:cohort_summary}
\resizebox{\textwidth}{!}{%
\begin{tabular}{lcccc}
\toprule
\textbf{Tier} & \textbf{Participants} & \textbf{SNP variants (raw)} & \textbf{SNP variants (post-QC, approx.)} & \textbf{ECG features (QC)} \\
\midrule
Tier 1 & 350  & \multirow{3}{*}{33,000,000 (dataset-level)} &  \(\approx\) 387,000 & 12 \\
Tier 2 & 500  &  & \(\approx\) 554,000 & 12 \\
Tier 3 & 6,006 &  & \(\approx\) 6,659,000 & 12 \\
\midrule
Total  & 8,856 & 33,000,000 & \(\approx\) 9,800,000 & 12 \\
\bottomrule
\end{tabular}%
}
\end{table*}

Only $350$ participants have confirmed cardiac diagnoses, so we stratified the cohort into three clinically motivated tiers to support pseudo-label generation and fine-tuning~\ref{fig:example-generation}:

\begin{itemize}
\item \textbf{Tier 1 ($L_1$):} Participants with high-confidence cardiac diagnoses, such as atrial fibrillation or myocarditis.
\item \textbf{Tier 2 ($L_2$):} Participants exhibiting indirect cardiovascular risk factors, including hypertension or hyperlipidemia.
\item \textbf{Tier 3 ($L_3$):} Unlabeled participants with no known prior cardiac diagnosis.
\end{itemize}

Participants were assigned to tiers using a combination of keyword matching and similarity-based strategies with BioBERT embeddings~\cite{deka2022evidence}. Tier 1 captures individuals with well-established, high-risk conditions, Tier 2 includes those with indirect phenotypes associated with cardiovascular risk, and Tier 3 encompasses the remaining participants without known diagnoses. This stratification provides a structured framework for pseudo-label generation and Top-k cluster selection in the training pipeline.

\section{Model Training Pipeline}
\subsection{Stage 1: Pseudo-Label Generation}

We construct pseudo-labels across all three tiers. SNP and ECG features are first preprocessed by filtering stop words and irrelevant variants, which then allows mapping of SNP rsIDs to relevant conditions in Tier 1 using curated GWAS repositories~\cite{ebiGWASCatalog}, and ECG features are standardized to morphological and temporal parameters such as PR interval, QTc, and heart rate variability.

The cleaned feature sets are then embedded into dense vector representations. For SNP data, we utilize two complementary strategies: (i) curated disease-specific variant sets for Tier 1 participants, ensuring biologically grounded feature selection; and (ii) TF-IDF encoding for Tier 2 and Tier 3 participants, treating SNP rsIDs as tokens to highlight rare, informative variants~\cite{10876089}. ECG timeseries features are projected into embeddings using a transformer-based encoder. All embeddings are concatenated into a unified multimodal representation, which serves as input to the k-means clustering algorithm.

K-means is applied to derive latent genotype–phenotype groupings ($k=20$). Each participant is then assigned a pseudo-label corresponding to its cluster. The value of $k$ was chosen empirically based on the underlying genetic structure and diversity of cardiovascular-associated variants. Specifically, preliminary analyses of linkage disequilibrium (LD) patterns and allele frequency distributions across SNPs indicated the presence of approximately 18-22 distinct subgroups representing coherent genotype clusters. Selecting $k=20$ provided a biologically interpretable granularity that balances between overfragmentation (high $k$) and information loss (low $k$). This clustering resolution aligns with previous studies in genomic subtyping and population stratification~\cite{li2019genetic, manolio2020implementing}, where $k$ values between 10 and 25 have been shown to effectively capture genetic heterogeneity without overfitting noise.

\subsection{Stage 2: Top-k Cluster Selection and Fine-Tuning}

The generated pseudo-labeled data are split into training (75\%) and testing (25\%) subsets. A pretrained transformer (domain-adapted BioBERT model for medical text) is fine-tuned on the training subset of pseudo labels. Model performance on the held-out pseudo labels is evaluated using accuracy and semantic similarity.

The top-$k$ clusters are then selected based on predictive consistency, forming the refined pseudo-labeled dataset. This stage ensures that only the most reliable cluster-based labels contribute to downstream model optimization. The retained clusters are then used to further fine-tune the language model in the PL-FT (Pseudo-Label Fine-Tuning) component, augmenting its ability to reason about multimodal signals in a clinically meaningful way.

\subsection{Few-Label Fine-Tuning (Cls-FT)}

Finally, the model pretrained on pseudo labels (PL-FT) is adapted to a small set of gold-standard clinical labels. For evaluation, we stratified a balanced subset of $1,050$ participants ($350$ per tier). Using parameter-efficient fine-tuning (LoRA with rank $8$ and $\alpha = 16$), we further refine the model on these labels while preserving efficiency and avoiding overfitting.

\begin{figure*}[ht]
    \centering
    \includegraphics[width=0.89\textwidth]{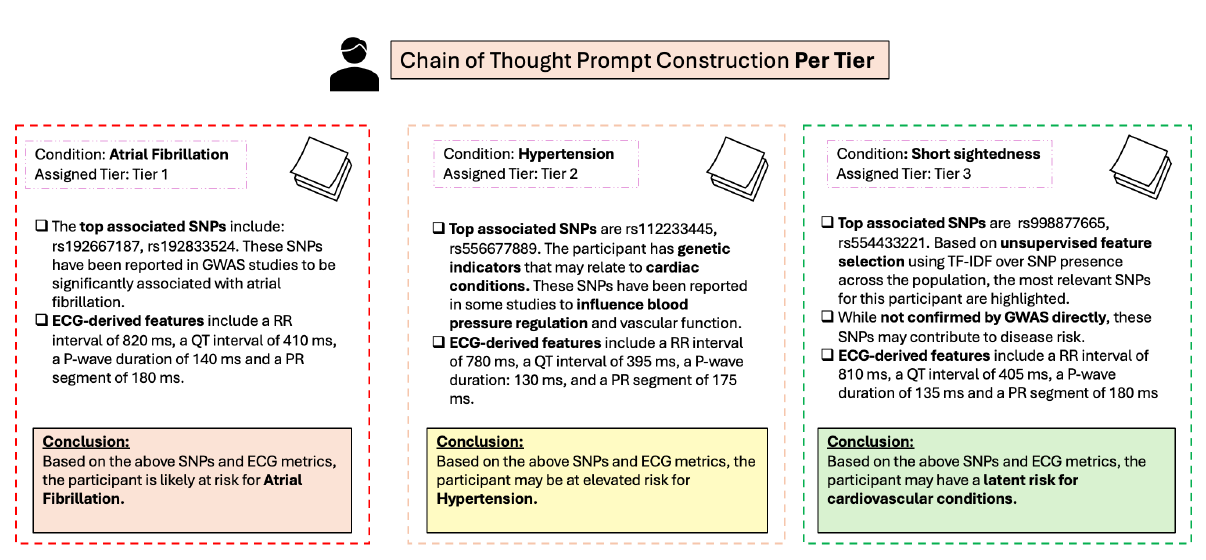}
    \caption{Example Creation of a Prompt for participants across tiers. The text in bold indicates the important features that the LLMs will choose to focus on, and the conclusions allow LLMs to draw the reasoning and link it to the relevant pseudo-label.}
    \label{fig:cot-prompt}
\end{figure*}

Chain-of-Thought (CoT) prompting~\cite{miao_chain_2024,wei2023chainofthoughtpromptingelicitsreasoning} is employed in this phase to guide reasoning. Each prompt integrates ECG-derived features, disease-associated SNPs, and tier-specific risk annotations, and concludes with an instructional query asking the model to infer potential cardiovascular risk, as showcased in tableo~\ref{fig:cot-prompt}. This structured prompting enforces stepwise reasoning and improves interpretability. 

The model outputs are evaluated semantically rather than by exact string match to outline for any similar keywords present. Predictions are embedded using BioBERT-based SentenceTransformers~\cite{deka2022evidence}, and cosine similarity is used to measure alignment with ground-truth labels. A high threshold of $0.7$ determines semantic correctness. This evaluation framework ensuring that the clinically relevant predictions are recognized even when lexical variations may occur.

\section{Results and Discussion}

This section presents the evaluation of the three selected base causal LLMs: GPT-2~\cite{radford2019language}, DeepSeek 1.3B~\cite{guo_deepseek-coder_2024}, and Llama 3.2 1B~\cite{noauthor_llama_2025} on the few-label, multimodal cardiovascular risk prediction task. Models were fine-tuned using LoRA-based parameter-efficient fine-tuning (PEFT)~\cite{peft}, with generation restricted to $512$ tokens per response. A semantic similarity threshold of $0.7$ was applied to ensure generated outputs were closely aligned with ground-truth labels, mitigating unreliable predictions.

\subsection{Performance}

Model performance was assessed using accuracy, precision, recall, F1-score, and evaluation loss metrics, both across the entire cohort and stratified tiers. Tables~\ref{tab:model_performance_overall}–\ref{tab:model_performance_tier3} summarize these results. 
\newline
The overall performance reflects model accuracy and generalization across all participants. Tier 1 consists of participants with high-confidence, clinically confirmed cardiac conditions, providing a clear ground-truth signal. Tier 2 includes participants with indirect or secondary cardiovascular risk factors (e.g., hypertension or hyperlipidemia), representing moderately challenging cases where phenotypic links are weaker. Tier 3 contains participants without confirmed diagnoses, serving as a test of the model's ability to infer risk and latent genotype–phenotype associations from unsupervised or pseudo-labeled data. This tiered evaluation allows us to examine performance across varying levels of label availability and difficulty, providing a comprehensive understanding of model behavior.

To evaluate the contribution of pseudo-labeling and multimodal integration, an ablation study was conducted. Specifically, we compared three configurations for each model: (i) the baseline model trained on the combined SNP–ECG feature set with pseudo-labels, (ii) a variant trained without ECG inputs (No ECG), and (iii) a variant trained without SNP inputs (No SNP). This setup allows us to quantify the value of multimodal fusion and the degree to which each modality contributes to downstream prediction accuracy. 

\begin{table}[ht]
\centering
\setlength{\tabcolsep}{4pt} 
\renewcommand{\arraystretch}{0.90}
\caption{Ablation study comparing model performance  under a multimodal (SNP + ECG) versus a unimodal (genotype only or phenotype only) configuration. The best results for each metric and the overall best model across all LLMs are in bold.}
\label{tab:ablation_results}
\resizebox{\linewidth}{!}{
\begin{tabular}{lccccc}
\toprule
\textbf{Model} & \textbf{Type of Split} & \textbf{Accuracy} & \textbf{Precision} & \textbf{Recall} & \textbf{F1-Score} \\
\midrule
\multirow{3}{*}{GPT-2} 
 & Baseline & 0.810 & 0.822 & 0.840 & 0.830 \\
 & No ECG & \textbf{0.710} & \textbf{0.703} & \textbf{0.702} & \textbf{0.700} \\
 & No SNP & 0.740 & 0.754 & 0.740 & 0.745 \\
\midrule
\multirow{3}{*}{DeepSeek 1.3B}
 & Baseline & 0.920 & 0.831 & 0.810 & 0.820 \\
 & No ECG & \textbf{0.810} & \textbf{0.791} & \textbf{0.724} & \textbf{0.746} \\
 & No SNP & 0.822 & 0.711 & 0.723 & 0.714 \\
\midrule
\multirow{3}{*}{Llama 3.2 1B}
 & Baseline & 0.920 & 0.830 & 0.891 & 0.840 \\
 & No ECG & \textbf{0.702 }& \textbf{0.691} & \textbf{0.650} & \textbf{0.644} \\
 & No SNP & 0.755 & 0.722 & 0.724 & 0.726 \\
\bottomrule
\end{tabular}
}
\end{table}

As shown in Table~\ref{tab:ablation_results}, removing either modality leads to a significant performance drop across all metrics, confirming that the fusion of genomic and ECG features contributes synergistically to cardiovascular risk prediction. Notably, the models that do not consist of ECG inputs exhibited greater declines in recall, suggesting that dynamic cardiac signals play a key role in identifying subtle risk indicators that static SNP embeddings alone cannot capture. Conversely, removing SNPs led to reduced precision, indicating that genotype information enhances the model’s ability to differentiate true positives from confounders. These findings demonstrate the value of multimodal learning for few-label biomedical prediction tasks. 




\begin{table}[ht]
\centering
\caption{Overall performance comparison of LLMs under a few labels setting and skyline overview. The best results for each metric and the overall best model across all LLMs are in bold. }
\label{tab:model_performance_overall}
\resizebox{\linewidth}{!}{
\begin{tabular}{lcccc}
\toprule
\textbf{Model} & \textbf{Accuracy} & \textbf{Precision} & \textbf{Recall} & \textbf{F1} \\
\midrule
& &\textbf{ Few Labels [1050 participants]} & & \\ \hline
GPT-2         & 0.811 & \textbf{0.890 }& 0.812 & 0.842 \\
LLaMA-3.2 1B  & 0.880 & 0.822 & 0.790 & 0.790 \\
\textbf{DeepSeek 1.3B} & \textbf{0.892} & 0.860 & 0.840 & 0.832 \\
\midrule
& & \textbf{Skyline [8,856 participants]} & & \\\hline
GPT-2         & 0.800 & 0.810 & 0.809 & 0.810 \\
LLaMA-3.2 1B  & 0.901 & 0.832 & 0.780 & 0.790 \\
DeepSeek 1.3B & 0.910 & 0.869 & 0.830 & 0.840 \\
\bottomrule
\end{tabular}
}
\end{table}

\begin{table}[ht]
\centering
\caption{Performance for Tier 1 participants under a few labels setting and skyline overview. The best results for each metric and the overall best model across all LLMs are in bold.}
\label{tab:model_performance_tier1}
\resizebox{\linewidth}{!}{
\begin{tabular}{lcccc}
\toprule
\textbf{Model} & \textbf{Accuracy} & \textbf{Precision} & \textbf{Recall} & \textbf{F1} \\
\midrule
& & \textbf{Few Labels [350 participants]} & & \\ \hline
GPT-2         & 0.810 & 0.822 & 0.840 & 0.830 \\
LLaMA-3.2 1B    & \textbf{0.920} & 0.830 & \textbf{0.891} & \textbf{0.840} \\
\textbf{DeepSeek 1.3B} & \textbf{0.920} & \textbf{0.831} & 0.810 & 0.820 \\
\midrule
 & & \textbf{Skyline [Full Data]} & & \\\hline
GPT-2         & 0.820 & 0.825 & 0.845 & 0.835 \\
LLaMA-3.2 1B  & 0.925 & 0.840 & 0.895 & 0.855 \\
DeepSeek 1.3B & 0.935 & 0.870 & 0.835 & 0.845 \\
\bottomrule
\end{tabular}
}
\end{table}

\begin{table}[ht]
\centering
\caption{Performance comparison for Tier 2 participants under a few labels setting and skyline overview. The best results for each metric and the overall best model across all LLMs are in bold.}
\label{tab:model_performance_tier2}
\resizebox{\linewidth}{!}{
\begin{tabular}{lcccc}
\toprule
\textbf{Model} & \textbf{Accuracy} & \textbf{Precision} & \textbf{Recall} & \textbf{F1} \\
\midrule
& & \textbf{Few Labels [350 participants]} & & \\ \hline
GPT-2         & 0.800 & 0.813 & 0.791 & 0.800 \\
LLaMA-3.2 1B    & 0.890 & 0.824 & \textbf{0.820} & 0.822 \\
\textbf{DeepSeek 1.3B} & \textbf{0.910} & \textbf{0.850 }&\textbf{ 0.820 }& \textbf{0.830} \\
\midrule
 & & \textbf{Skyline [Full Data]} & & \\\hline
GPT-2         & 0.810 & 0.820 & 0.800 & 0.810 \\
LLaMA-3.2 1B  & 0.905 & 0.835 & 0.825 & 0.830 \\
DeepSeek 1.3B & 0.925 & 0.860 & 0.830 & 0.840 \\
\bottomrule
\end{tabular}
}
\end{table}

\begin{table}[ht]
\centering
\caption{Performance comparison for Tier 3 participants under a few labels setting and skyline overview. The best results for each metric and the overall best model across all LLMs are in bold.}
\label{tab:model_performance_tier3}
\resizebox{\linewidth}{!}{
\begin{tabular}{lcccc}
\toprule
\textbf{Model} & \textbf{Accuracy} & \textbf{Precision} & \textbf{Recall} & \textbf{F1} \\
\midrule
& & \textbf{Few Labels [350 participants]} & & \\ \hline
GPT-2         & 0.811 & \textbf{0.890 }& 0.812 & \textbf{0.842} \\
LLaMA-3.2 1B    & 0.880 & 0.822 & 0.790 & 0.790 \\
\textbf{DeepSeek 1.3B } & \textbf{0.892 }& 0.860 & \textbf{0.840} & 0.832 \\ \hline
\midrule
  & & \textbf{Skyline [Full Data]} & & \\\hline
GPT-2         & 0.820 & 0.895 & 0.815 & 0.850 \\
LLaMA-3.2 1B  & 0.890 & 0.830 & 0.795 & 0.805 \\
DeepSeek 1.3B & 0.900 & 0.870 & 0.845 & 0.855 \\
\bottomrule
\end{tabular}
}
\end{table}

Following the ablation study, we next evaluate the end-to-end models on the genotype–phenotype relation task. This evaluation isolates the model’s ability to infer cardiovascular risk from integrated genetic and electrophysiological signals. By comparing across model architectures, we assess not only predictive strength but also how pretraining scale and multimodal reasoning capacity affect generalization under few-label supervision. Here, DeepSeek 1.3B achieved the highest performance overall, with an accuracy of $0.910$, precision of $0.869$, recall of $0.830$, and F1 score of $0.840$. This suggests that DeepSeek is the most effective in correctly classifying samples while maintaining a balanced trade-off between precision and recall. This is followed by Llama 3.2 1B that achieves a $0.901$ accuracy and a moderately lower recall of $0.780$, which slightly lowered its F1 score to $0.790$. Finally, GPT-2 performs reasonably well with a F1 score of $0.810$, but is lagged behind the more recent and larger models.

Collectively, these results highlight how model scale and pretraining diversity influence adaptation efficiency. DeepSeek’s performance advantage suggests that pretraining on diverse biomedical and technical corpora enhances its capacity to interpret multimodal representations. Conversely, GPT-2’s slower convergence underscores the limitations of smaller, older architectures in capturing complex genotype–phenotype dependencies.

\begin{figure}[ht]
    \centering
    \includegraphics[width=0.35\textwidth]{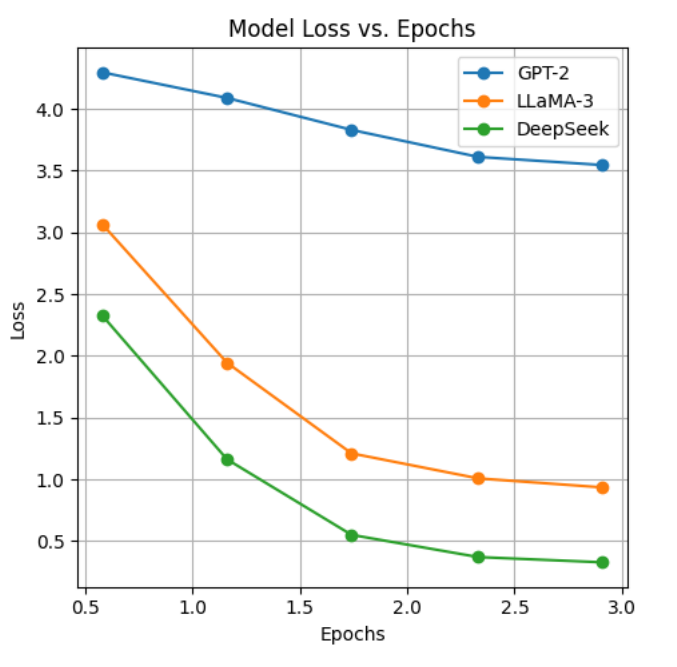}
    \caption{Epoch-wise training loss comparison accross three models.}
    \label{fig:loss-per-training-step}
\end{figure}

Furthermore, the figure \ref{fig:loss-per-training-step} compares the training loss curves of these fine-tuned models across approximately 3 epochs. The DeepSeek fine-tuning exhibits the most rapid convergence, achieving a loss below $0.5$ by epoch $\approx1.75$. In contrast, Llama 3.2 1B shows moderate convergence, while GPT-2 converges the slowest, with a final loss of $3.546$, consistent with the nature of the models that has been previously exhibited.

This trend suggests that model architecture and pretraining quality significantly influence the efficiency of adaptation in fine-tuning. DeepSeek’s fast convergence may indicate better initialization or pretraining alignment with the task data. Meanwhile, GPT-2, being a smaller and older architecture, shows slower adaptation and likely requires either more epochs or lower learning rates to match performance.

Additionally, the shape of each curve provides insight into optimization stability. All three models exhibit monotonic decreases in loss, indicating stable training with no signs of overfitting within the observed window. However, DeepSeek’s steep initial drop followed by a plateau suggests that it reaches saturation quicker, while GPT-2 still shows room for improvement beyond 3 epochs.

\subsection{Tier-wise Analysis}

To investigate model behavior across data with varying label density and difficulty, we performed tier-specific analyses:

\subsubsection{Tier 1 Participants}

Tier 1 participants included those with confirmed cardiac diagnoses and well-established genotype-phenotype associations. DeepSeek 1.3B and Llama 3.2 1B achieved equal accuracy ($0.920$). Llama 3.2 1B exhibited higher recall ($0.891$ vs. $0.810$), capturing more true positives, while DeepSeek 1.3B demonstrated balanced precision and recall, yielding comparable F1 ($0.820$). GPT-2 performed slightly lower overall (F1 $0.830$).

\subsubsection{Tier 2 Participants}

The classification of Tier 2 participants exhibited indirect cardiovascular risk factors. DeepSeek 1.3B led with $0.910$ accuracy and F1 $0.830$, showing robustness to intermediate phenotypes. Llama 3.2 1B followed closely, while GPT-2 remained consistent but lower (F1 $0.800$). These results indicate that larger, newer LLMs better generalize to moderately challenging, partially labeled datasets.

\subsubsection{Tier 3 Participants}

Tier 3 participants lacked direct disease annotations, requiring inference from ECG morphology and clustered genotype features. This is why DeepSeek 1.3B performed considerably lower than the previous participants, achieving an F1 $0.832$ with balanced precision ($0.86$) and recall ($0.84$), while still outperforming other models. Llama 3.2 1B also showed lower recall ($0.79$), while GPT-2 surprisingly performed slightly better than in Tier 2, likely reflecting overfitting to simple latent patterns within the pseudo-labeled clusters.

\subsection{Skyline vs Few-Labels Performance}

To better understand the impact of supervision density, we compared each model’s performance under full supervision (Skyline) versus the few-label setting (350 labels). Across all tiers, the Skyline configuration consistently achieved marginally higher accuracy and F1 scores, confirming its role as an upper bound on achievable performance. However, the relative gap between Skyline and Few-Labels was notably small for DeepSeek 1.3B—often less than $0.02$ in F1—indicating strong generalization even with limited supervision. This contrasts with LLaMA-3.2 1B and GPT-2, which exhibited larger drops in recall under few-label conditions, suggesting greater sensitivity to sparse supervision.

We found that the few-label setting occasionally produced slightly higher precision in GPT-2 and LLaMA models, a phenomenon also observed in semi-supervised learning literature. This can occur when limited supervision biases the model toward high-confidence predictions, improving precision at the cost of recall. DeepSeek’s consistent performance across both settings implies that its pretraining diversity and architecture enable robust embedding alignment, allowing it to leverage pseudo-labels and latent structure more effectively than smaller models.

To further alleviate these results, we decided to compare model performance for participants across different sizes of labels.

\begin{table}[ht]
\centering
\caption{Performance comparison of LLMs on participants with 50–350 labels.}
\label{tab:model_performance_across_labels}
\resizebox{0.8\linewidth}{!}{
\begin{tabular}{lcccc}
\toprule
\textbf{Model} & \textbf{Accuracy} & \textbf{Precision} & \textbf{Recall} & \textbf{F1} \\
\hline
\multicolumn{5}{l}{\textbf{50 labels}} \\
\hline
DeepSeek & 0.60 & 0.62 & 0.58 & 0.60  \\
LLaMA & 0.59  & 0.60 & 0.58 & 0.59  \\
GPT-2 & 0.62  & 0.61 & 0.63 & 0.62  \\
\hline
\multicolumn{5}{l}{\textbf{100 labels}} \\
\hline
DeepSeek & 0.63  & 0.64 & 0.62 & 0.63  \\
LLaMA & 0.62  & 0.63 & 0.61 & 0.62  \\
GPT-2 & 0.58  & 0.57 & 0.59 & 0.58  \\
\hline
\multicolumn{5}{l}{\textbf{150 labels}} \\
\hline
DeepSeek & 0.65  & 0.66 & 0.64 & 0.65  \\
LLaMA & 0.67  & 0.68 & 0.66 & 0.67  \\
GPT-2 & 0.64  & 0.63 & 0.65 & 0.64  \\
\hline
\multicolumn{5}{l}{\textbf{200 labels}} \\
\hline
DeepSeek & 0.70  & 0.71 & 0.69 & 0.70  \\
LLaMA & 0.68  & 0.69 & 0.67 & 0.68  \\
GPT-2 & 0.68  & 0.67 & 0.69 & 0.68  \\
\hline
\multicolumn{5}{l}{\textbf{250 labels}} \\
\hline
DeepSeek & 0.76  & 0.77 & 0.75 & 0.76  \\
LLaMA & 0.723 & 0.73 & 0.72 & 0.72  \\
GPT-2 & 0.74  & 0.73 & 0.75 & 0.74  \\
\hline
\multicolumn{5}{l}{\textbf{300 labels}} \\
\hline
DeepSeek & 0.83  & 0.84 & 0.82 & 0.83  \\
LLaMA & 0.82  & 0.83 & 0.81 & 0.82  \\
GPT-2 & 0.78  & 0.77 & 0.79 & 0.78  \\
\hline
\multicolumn{5}{l}{\textbf{350 labels}} \\
\hline
DeepSeek & 0.91  & 0.91 & 0.90 & 0.91  \\
LLaMA & 0.90  & 0.90 & 0.89 & 0.90  \\
GPT-2 & 0.80  & 0.81 & 0.79 & 0.80  \\
\bottomrule
\end{tabular}
}
\end{table}

As shown in Table~\ref{tab:model_performance_across_labels}, DeepSeek 1.3B, Llama 3.2 1B, and GPT-2 all exhibit improvements in accuracy and F1 score as the number of labels increases. Notably, DeepSeek and Llama initially show relatively lower performance with very few labels (50–100), likely due to overfitting tendencies when exposed to extremely sparse supervision. As additional labeled data become available, these larger models leverage their representational capacity to rapidly improve, with DeepSeek achieving the highest F1 of $0.91$ at 350 labels. GPT-2, being smaller, shows more modest gains and plateaus earlier, reflecting its limited capacity to capture complex multimodal genotype–phenotype relationships.

Overall, the small skyline and few-Labels gap demonstrates that with sufficient architectural and representational capacity, large language models can achieve near-skyline performance using only a fraction of labeled data. However, when the number of labels is extremely limited, larger models such as DeepSeek 1.3B and Llama 3.2 1B can initially underperform due to overfitting tendencies on sparse supervision. As more labeled samples are introduced, their performance rapidly improves, highlighting the importance of label quantity in stabilizing training while still taking advantage of the models’ representational capacity. This finding strongly supports the scalability of few-label paradigms and the need for careful monitoring of overfitting in biomedical contexts where annotation is costly or infeasible.

\subsection{Discussion}

DeepSeek 1.3B consistently outperformed other models across tiers, demonstrating effective learning from sparse, pseudo-labeled data. This indicates that DeepSeek's pretraining scale and architechture have advantages in embedding high-dimenstional features, particularily when supervision is limited or weak. Its superior between recall and precision suggests it identifies the high-confidence association while also generalizing beyond labeled examples, indicative of effective few-label learning. Llama 3.2 1B was competitive but occasionally struggled with recall in cases requiring latent inference from multimodal inputs. GPT-2, though limited in scale, provided a strong baseline, confirming that even smaller architectures can capture key genotype–phenotype relations under few-label supervision. These patterns suggest that DeepSeek’s larger multimodal embedding space allows more effective genotype–phenotype reasoning, while Llama’s recall gap may indicate suboptimal adaptation to ECG signals. The stable baseline of GPT-2 highlights the benefit of even modest LLMs for structured biomedical reasoning.

Training loss progression highlighted DeepSeek’s rapid convergence, reflecting its capacity to efficiently integrate few-label multimodal signals. Llama 3.2 1B converged more slowly, while GPT-2 exhibited higher loss and slower adaptation. This suggests that model architecture and representational capacity play a critical role in few-label cardiovascular risk prediction.

The tiered evaluation demonstrates the importance of assessing model performance across varying label availability. Few-label strategies, combined with chain-of-thought (CoT) reasoning and pseudo-labeling, allow LLMs to generalize from limited supervision while producing interpretable outputs relevant to clinical risk assessment. Notably, Tier 3 results indicate that pseudo-label propagation can capture latent genotype–phenotype associations even in the absence of explicit diagnostic labels, underscoring the framework’s potential for semi-supervised discovery in real-world biomedical datasets. In addition, the skyline versus few-labels comparison further reinforces that model scale and pretraining diversity modulate label efficiency. DeepSeek’s narrow performance gap between supervision regimes highlights its adaptability to low-resource conditions—a critical property for translational use in biomedical contexts where labeled data remain scarce.

Analyzing performance across increasing label counts (50–350 labels) further illustrates the impact of supervision density on model learning. DeepSeek 1.3B consistently improves with more labels, achieving near-perfect accuracy and balanced precision/recall at 350 labels, while Llama 3.2 1B shows steady gains and GPT-2 plateaus earlier. These results indicate that models with larger multimodal embedding spaces and diverse pretraining are better able to capitalize on additional labeled data, whereas smaller architectures benefit less from incremental supervision. The trends follows closely with the Skyline vs Few-Labels findings, emphasizing that effective few-label strategies, particularly pseudo-labeling combined with PEFT, allow large LLMs to approach full-data performance with substantially fewer annotations.

From a biological perspective, the integration of SNP and ECG-derived representations enables the model to attend to both static genetic risk markers and dynamic physiological patterns. The resulting multimodal embeddings facilitate deeper understanding of the molecular and electrophysiological mechanisms underlying cardiovascular disease. This synergy between modalities aligns with emerging evidence that complex cardiac risk phenotypes are jointly shaped by genomic and electrical factors rather than isolated signals. The joint SNP-ECG embeddings may therefore reflect an emergent representation of cardiac physiology, where genomic variants shape electrophysiological expression patterns observable through ECG signals. This alignment between model attention and biological mechanism highlights the framework's potential for discovery, not just prediction.

However, several limitations should be noted. The present framework is evaluated on a subset of SNP and ECG features, which may not capture the full diversity of cardiac phenotypes. Additionally, the pseudo-labeling strategy, while effective, may propagate noise in sparsely annotated regions. Future work could incorporate confidence-weighted labeling or uncertainty calibration to mitigate this. Scaling to larger models (e.g., Llama-3 7B or GPT-4) and integrating additional modalities such as imaging, clinical notes, or biochemical data may further enhance predictive robustness and clinical interpretability. Despite these limitations, the proposed few-labels approach represents a viable route towards practical deployment in a low-resource clinical setting, where annotated multimodal data remains scarce but unlabeled signals are abundant. 

Overall, these findings highlight that parameter-efficient LLMs, when combined with tiered pseudo-labeling and CoT reasoning, offer a promising direction for interpretable, multimodal cardiovascular risk modeling. This framework provides a scalable foundation for next-generation clinical decision-support systems that balance predictive performance with transparency, a critical requirement for AI deployment in medicine.

\section*{Conclusion}

In this study, we present a few-label multimodal LLM framework for interpretable cardiovascular risk stratification, demonstrating that effective integration of SNP genotypes and ECG-derived phenotypes can uncover meaningful genotype-phenotype relationships even in data-scarce environments. By employing a three-tier pseudo-labeling system, the framework accommodates varying levels of annotation certainty and sparsity, enabling models to learn both direct and latent genotype–phenotype relationships. Chain-of-Thought (CoT) prompting further enhances transparency by producing structured clinical rationales alongside model predictions.

Employing an empirical evaluation across three large language model architectures (GPT-2, Llama 3.2 1B, and DeepSeek 1.3B) reveals several key findings. Firstly, DeepSeek 1.3B consistently achieves the highest overall and tiered performance, underscoring the benefit of large-scale pretraining in few-label biomedical settings. The ablation study confirms that removing either genomic or ECG inputs leads to notable performance degradation, demonstrating that multimodal fusion contributes synergistically to cardiovascular risk prediction. Moreover, the pseudo-labeling system effectively extends model generalization to under-annotated cohorts—particularly Tier~3 participants—while maintaining clinically interpretable reasoning outputs.

The interpretability analyses show that the CoT prompts guide models toward biologically grounded explanations, highlighting how models reference relevant SNP loci and ECG waveform characteristics to justify predictions and offering interpretable pathways that align with known cardiovascular mechanisms. This capacity for transparent inference is a crucial step toward clinical trustworthiness and model accountability.

Despite these promising outcomes, there are still some certain limitations that remain to be addressed. The current methodology focuses primarily on genomic and ECG modalities. However, future work could incorporate additional data sources such as imaging, laboratory, and lifestyle variables to enhance both predictive accuracy and biological resolution. Also, scaling the framework to larger architectures (e.g., Llama-3 7B~\cite{llama3modelcard} or GPT-4~\cite{gpt4}) and integrating structured biomedical knowledge bases~\cite{xu_knowledge_2024} may further improve reasoning over complex genotype–phenotype–disease pathways. Additionally, extending the pseudo-labeling approach to semi-supervised or contrastive settings could enhance robustness to label noise and cohort heterogeneity. 

Overall, this study demonstrates that parameter-efficient LLMs, when coupled with tiered pseudo-labeling and Chain-of-Thought reasoning, offer a scalable and interpretable framework for multimodal cardiovascular prediction. By unifying genomic and physiological representations, the proposed method achieves both strong predictive performance and clinically meaningful transparency—advancing the development of few-label, explainable AI systems for precision cardiovascular medicine and beyond. Ultimately, this work underscores how parameter-efficient LLMs can serve as an adaptable foundation for future biomedical applications that demand both accuracy and explainability, which serves as a bridge in the gap between computational prediction and clinical interpretation. 

\bibliographystyle{IEEEtran}
\bibliography{acl2016}
\printglossaries

\end{document}